# Predicting credit default probabilities using Bayesian Statistics and Monte-Carlo Simulations

Dominic Joseph

**Abstract**

Banks and financial institutions all over the world manage portfolios containing tens of thousands of customers. Not all customers are high credit-worthy, and many possess varying degrees of risk to the Bank or financial institutions that lend money to these customers. Hence assessment of credit risk is paramount in the field of credit risk management. This paper discusses the use of Bayesian principles and simulation-techniques to estimate and calibrate the default probability of credit ratings. The methodology is a two-phase approach where, in the first phase, a posterior density of default rate parameter is estimated based the default history data. In the second phase of the approach, an estimate of true default rate parameter is obtained through simulations.

## 1. Introduction

Probability of default (PD) or Default Rate is a financial risk management term describing the likelihood of a default over a particular time horizon. It provides an estimate of the likelihood that a borrower will be unable to meet its debt obligations. Under Basel II guidelines, formulated by the Basel Committee on Banking Supervision or BCBS (2001), PD is a key parameter used in the calculation of economic capital or regulatory capital for a banking institution. Banks that comply with the new Basel II internal ratings-based approach are obliged to assign a PD value (usually a 1-year PD) to their clients, which are used as the basis for regulatory capital requirements.

A popular method of quantifying Probability of default is through credit ratings, where each entity in a bank's portfolio is assigned a rating grade depending on its past, current and future behaviour. Entities with same rating grade represents similar credit risk. Some banks have internal score card and rating models that assign ratings based on internal methodologies. On occasions, banks may use ratings supplied by external credit rating agencies such as Moody's and Standard & Poor's (S&P) instead of internal ratings. For example, S&P Global ratings provide a rating system with 22 rating classes and Moody's, which is another popular rating agency classifies entities to 21 rating classes based on their credit worthiness. This is permissible because Basel II accord allows banks to base their capital requirements on internal as well as external rating systems and risk profile.

Large banks employ various mathematical techniques using the historical data of customer or account level ratings assigned by its internal ratings system (or external rating system) to forecast the default probability of each entity. The number of grades used by the bank depends on the bank's individual preference. The "upper" grades represent lower default risk and hence lower probability values are assigned to them while "lower" grades represent higher default risk and consequently, higher probability values are assigned to them.

One of the important concerns of analysts in the financial industry is that the empirically estimated PD values of rating classes are non-monotonic in nature (Katja Pluto and Dirk Tasche, 2005), i.e. $R' > R$ ($R'$ is "lower" grade) need not imply $\hat{\theta}_{R'} > \hat{\theta}_R$. Where $\hat{\theta}_R$ is the estimated default rate for rating grade R. Dirk Tasche (2013) discussed this issue which he termed it as the 'Inversion of default rates'. Non monotonicity is an issue because the observed default rates do not reflect the true default risk associated with the rating class. Krahnen, J.P & Weber, M. (2001) mentions monotonicity as one of the most important requirements for a rating system. It is easy to see that without monotonicity, the situation can lead to issues like misclassification of some customers into wrong risk buckets with under-estimation or over-estimation of default probability. It also affects the discriminatory power of a rating system thereby hindering meaningful risk differentiation between less risk and high-risk rating classes.

One reason for break in monotonic trend could be that the portfolio of most banks has an uneven distribution of accounts in the rating classes (Dirk Tasche, 2013). It is commonly observed by industry experts in many of the data that are analysed, that there is insufficient data in some rating groups, thereby making it impossible to obtain an accurate estimate of its default risk from the observed default rate itself. For example, Table 1 shows the observed default rates based on historical rating assignments (Long-Term Foreign-Currency) of corporate assets by Standard & Poor's Ratings Services for the years 2017 and 2016. Note that for both years, there is a break in monotonicity in the empirically obtained default rates. Such inversions need not be a problem with the rank-ordering capacity of the rating methodology. The issuer weighted long-run average grade-level default rates reported by S&P (2020, table 4) suggests that the observation of default rate inversions as in table 1 might be an exception.

*Table 1: S&P default frequencies and default rates (%) for corporate entities in 2016 and 2017 based on Long Term Foreign-Currency issuer rating*

| Rating | 2016 | | 2017 | |
| --- | --- | --- | --- | --- |
| | Default Frequency | Default Rate | Default Frequency | Default Rate |
| AAA | 0 | 0.0% | 0 | 0.0% |
| AA | 0 | 0.0% | 0 | 0.0% |
| A | 0 | 0.0% | 0 | 0.0% |
| BBB | 0 | 0.0% | 0 | 0.0% |
| BB | 60 | 4.1% | 2 | 0.1% |
| B | 25 | 2.0% | 14 | 0.5% |
| CCC | 38 | 11.6% | 31 | 8.3% |
| CC | 1 | 3.4% | 4 | 14.3% |
| C/D | Default Bucket | | | |

In such cases, it is safe to assume that the observed default rate is different from the true default rate. A lack of sufficient data might lead to a difficulty in calculating the true probability of default value for such rating grades. Hence there is a need to calibrate the estimated default probabilities such that,

$$R' > R \rightarrow \hat{\theta}_{R'} > \hat{\theta}_R$$

Some of the most popular techniques used in the industry are listed below:

i. Katja Pluto & Dirk Tasche (2005) suggested a low-default portfolio calibration approach, termed as the *most prudent estimation*, using upper confidence bound with confidence level (1-α), while guaranteeing an ordering of PDs that respects the differences in credit quality as indicated by the rating grades. This

  method involved solving an inequality by assuming a binomial likelihood of defaults and a suitable value for α

ii. Van der Burgt (2007) suggested a method for calibrating low-default portfolio PD curves by using the cumulative accuracy profile (CAP), also known as the power curve or Lorentz curve, and a mathematical function for modelling the CAP. This is also called *CAP Curve Calibration* or *VDB Calibration* technique.

iii. Dirk Tasche (2009) proposed a two-parameter approach called *Quasi Moment Matching* (QMM) based on Accuracy Ratio and distribution of good accounts. Here, the PD or default probability for each rating is modelled as a mathematical function of the distribution of non-defaulted population in the portfolio.

A recent work in the field using Bayesian framework was by Denis Surzhko (2017) who proposed a method where the portfolio default rate itself rather than rating level is calibrated to the default rate of another portfolio called the *closest available portfolio* by adjusting the Bayesian prior density of the default rate parameter. Although the method can be adopted for a rating level system, it involves identifying a closely related rated portfolio with reliable default statistics.

This paper proposes a combination of Bayesian approach and Monte-Carlo simulations for achieving the default rate curve calibration. Using Bayesian principles, we attempt to model the default parameter for each rating class R and period t given data as a continuous random variable $\theta_R(t)$ subject to certain conditions. We later see that the marginal density function of rating classes $\theta_R$ subject to a set of given restrictions, is an expression without any closed form solution. Hence a simulation approach is needed to be adopted for obtaining an approximate solution. The theoretical framework for the entire process has been discussed in detail in the following sections. The proposed methodology is modelled using S&P corporate default data and the results are presented in Section 3.3. The results are then compared to the outputs from other well-known approaches, which have been provided in Section 4.

## 2. Methodology

### 2.1. Beta Binomial Distribution

The beta-binomial distribution is the binomial distribution in which the probability of success at each of n trials is not fixed but randomly drawn from a beta distribution. Binomial distribution is the distribution of the number of successes that occur in n independent trials with the probability of success in each trial is $\theta$. If we consider n to be the number of performing entities at the beginning and the k denote the number of entities at the end who defaulted, then the distribution of number of defaults can be considered as a binomial distributed random variable with likelihood.

$$p(\theta, n) = \binom{n}{k} \theta^k (1-\theta)^{n-k} \qquad (1)$$

where $\theta$ is the true probability of occurrence of a default event of a single entity.

In the Bayesian framework, the parameter $\theta$ is assumed to have a probability distribution of its own called the prior. The suitable prior density for a binomial likelihood is a Beta-density.

$$\theta \sim Beta(\alpha, \beta) = \frac{\Gamma(\alpha + \beta)}{\Gamma(\alpha) \cdot \Gamma(\beta)} \cdot \theta^{\alpha-1}(1-\theta)^{\beta-1} \quad (2)$$

Beta distribution as a prior has the advantage that it is a conjugate prior to (1) and, therefore, allow us effective and simple posterior mean estimation. Beta posterior mean can be regressed on macro-variables (Ferrari and Cribari-Neto (2004)), so the model can be used seamlessly for stress-testing purposes.

## 2.2. Conditional Posterior Density

Bayesian Statistics is derived from Bayes theorem that describes the probability of an event, based on prior knowledge of conditions that might be related to the event. Bayesian statistics has three key components:

- A probability density called prior distribution ($Prior(\theta_R)$) shows the distribution of the parameter ($\theta_R$) before observing the evidence or data.
- Likelihood of data given $\theta_R$ ($Likelihood(x|\theta_R)$), which is the distribution of data $x = \{x_1, x_2 \ldots, x_n\}$, given the value of parameter $\theta_R$.
- After observing data we update our beliefs about the parameter $\theta_R$ given the knowledge of the data and prior knowledge of $\theta_R$ and calculate the posterior distribution ($Posterior(\theta_R|x)$).

$$Posterior(\theta_R|x) \; \alpha \; Likelihood(x|\theta_R) \times Prior(\theta_R) \quad (3)$$

In our case, the likelihood is the binomial likelihood discussed before.

$$Likelihood(x|\theta_R) = Binom(n_R, d_R) \quad (4)$$

At the beginning, the parameter $\theta$ is assumed to be uniformly distributed from 0 to 1, which is chosen as the prior distribution. This specification treats the PD as being unknown and assumes an uninformed or objective prior distribution. Therefore, it is assumed that there is no knowledge of what the realised PD might be.

$$\theta_R \sim U(0,1) = Beta(1,1) \quad (5)$$

Hence, Posterior density of $\theta_R$ comes out to be,

$$Posterior(\theta_R|x) = Binom(n_R, d_R) \times Beta(1,1) = Beta(1 + d_R, 1 + n_R - d_R) \quad (6)$$

For a portfolio with m ratings, the joint density of $(\theta_1, \theta_2, \ldots \theta_m)$ given the data is the product of each of its beta distribution density.

$$f(\theta_1, \theta_2, \ldots, \theta_m) = f(\theta_1) \times f(\theta_2) \times \ldots f(\theta_m) = \prod f(\theta_i) \quad (7)$$

This follows from the assumption that default event in one rating bucket doesn't affect the default event in any other rating bucket, i.e. $\theta_i$ and $\theta_j$ are independent random variables for $i \neq j$.

## 2.3. Calibration using Simulation

We know that the true value of the parameter follows the order

$$0 \leq \theta_1 \leq \theta_2 \leq \theta_3 \leq \cdots \leq \theta_m \leq 1$$

We assume that the reason that the observed value of the parameter breaks the order due to data limitations discussed earlier. Hence, we aim to solve for each $\theta_i$ under the constraint given by equation.

The marginal distribution of $\theta_i$ can be obtained from the joint density function as follows,

$$f(\theta_i | \theta_1 \leq \theta_2 \leq \cdots \leq \theta_m) = \frac{\int_{\theta_m=0}^{1} \cdots \int_{\theta_2=0}^{\theta_3} \int_{\theta_1=0}^{\theta_2} f(\theta_1, \theta_2, \ldots, \theta_m) \, d\theta_1 d\theta_2 \ldots d\theta_m}{P(\theta_1 \leq \theta_2 \leq \cdots \leq \theta_m)} \qquad (8)$$

This expression has no bounded solution. Hence, one can use a Monte-Carlo simulation approach to estimate the conditional marginal distribution of $f(\theta_i)$. Suppose $\{\theta'_{1i}, \theta'_{2i}, \theta'_{3i} \ldots \theta'_{Ni}\}$ is the set of N simulated random variables of $\theta_i$ from the conditional marginal distribution of $f(\theta_i)$.

If we assume that $\theta_i | \theta_1 \leq \theta_2 \leq \cdots \leq \theta_m \sim Beta(\alpha, \beta)$, then from the simulated numbers, method of moment matching can used to estimate the parameters $(\alpha, \beta)$ of the density function of each $\theta_i$.

Let $\bar{x}_i$ and $\bar{s}_i$ be the mean and standard deviation of the simulated samples for each rating group i.

$$\bar{x}_i = \frac{\sum \theta_i'}{N_i} \qquad (9)$$

$$\bar{s}_i = \sqrt{\frac{\sum (\theta_i')^2}{N_i} - \bar{x}_i^2} \qquad (10)$$

Using the technique of moment matching for beta distribution, we obtain that

$$\hat{\alpha}_i = \bar{x}_i \times \left( \frac{\bar{x}_i(1 - \bar{x}_i)}{\bar{s}_i^2} - 1 \right) \qquad (11)$$

$$\hat{\beta}_i = (1 - \bar{x}_i) \times \left( \frac{\bar{x}_i(1 - \bar{x}_i)}{\bar{s}_i^2} - 1 \right) \qquad (12)$$

where $(\hat{\alpha}_i, \hat{\beta}_i)$ is the set of estimated parameters of the conditional marginal distribution $f(\theta_i | \theta_1 \leq \theta_2 \ldots \theta_m)$ from the simulated sample $\{\theta_i'\}$. Appendix provides proof for the derivation of equations 11 and 12.

The calibrated probability of default for rating i can hence be estimated as the beta distribution mean

$$\mu_{\theta i} = \frac{\hat{\alpha}_i}{\hat{\alpha}_i + \hat{\beta}_i} \qquad (13)$$

## 3. Experimental Design

### 3.1 Data

"Standard & Poor's Ratings Services" issues credit ratings for the debt of public and private companies, and other public borrowers such as governments and governmental entities. It rates customers into 22 ratings based on their performance, with AAA being the best rating and minimal chance of default and D being the worst performance rating which stands for Default. Intermediate ratings are offered at each level between AA and CCC (such as BBB+, BBB, and BBB−). For the purpose of the modelling, Long-Term issuer ratings (Foreign-Currency) data of corporate customers rated by S&P was obtained[1]. The data consisted of ratings for the years 2016 & 2017.

For the purpose of simplicity and overcoming data challenges like lack of enough defaults in some rating grades, a binning process[2] was carried out where in adjacent ratings were clubbed together to form a new rating group with 9 ratings. For example, ratings BBB+, BBB and BBB- were merged to form a new group BBB. Also, Ratings C and D were considered to be the default bucket with rating order 9.

December snapshots for the years 2016 and 2017 was analysed for the modelling. A 12-month window was considered as an ideal observation window because banks and financial institutions are usually interested in measuring probability of default over a 1-year time horizon. For each snapshot period, the number of performing entities in the beginning and the number of default cases (from the performing entities) at the end of the 12 month window were measured for each rating class separately. See Table 2 for details of obligor and default counts measured for each year.

*Table 2: S&P corporate credit rating (Long-Term Foreign-Currency) and corresponding numerical order*

| Rating | Rating Order | 2016 | | 2017 | |
|---|---|---|---|---|---|
| | | Defaults at End | Performing at Start | Defaults at End | Performing at Start |
| AAA | 1 | 0 | 14 | 0 | 10 |
| AA | 2 | 0 | 153 | 0 | 148 |
| A | 3 | 0 | 934 | 0 | 978 |
| BBB | 4 | 0 | 1814 | 0 | 1952 |
| BB | 5 | 60 | 1470 | 2 | 1569 |
| B | 6 | 25 | 1225 | 14 | 2716 |
| CCC | 7 | 38 | 329 | 31 | 372 |
| CC | 8 | 1 | 29 | 4 | 28 |
| C/D | 9 | **Default bucket** | | | |

### 3.2 Sampling Algorithm

The sampling and calibration process is performed in an iterative and step by step manner.

i. Simulate n values of $\theta_i$ and $\theta_{i+1}$ (using estimated posterior distribution given by equation 6) for ratings i and i+1.
ii. Filter the simulated numbers such that $\theta_i \leq \theta_{1+1}$.

---

[1] Under SEC Regulation 17g-7, Nationally Recognized Statistical Rating Organizations (NRSRSOs) are required to report their historical rating assignments, upgrades, downgrades and withdrawals since 2010. Rating data are generally reported on a one year delay. CSV format of this data was sourced from the website www.ratingshistory.info.

[2] Note that this step is irrelevant from the point of view of actual modelling process and was chosen only for the purpose of convenience.

iii. For the filtered data, beta distribution parameters are estimated using moment matching technique, as explained earlier in section 2.3.
iv. Steps i to iii are continued till all the ratings are calibrated, with each step yielding new values for the beta distribution shape parameters. For each sampling process, updated beta distribution parameters from the previous sampling step is used.

The sampling processes[3] in steps i to iv is performed multiple times, to obtain a sampling distribution for each calibrated $\theta_i$. Once we have the sampling distribution, we can calculate mean, median and confidence intervals for our estimate of $\theta_i$.

## 3.3 Calibrated PD Output

Using the method of moment matching, Beta distribution parameters $(\hat{\alpha}_i, \hat{\beta}_i)$ are determined from the simulated random variables subject to given conditions explained earlier for each rating in each snapshot period. The estimated parameters for sample size n = 100,000 and n = 500,000 are as shown in Table 3 and Table 4, along with 90% confidence bounds.

*Table 3: Calibrated PD values for 100,000 simulations and 300 iterations*

| Simulated Results n = 100,000 K = 300 | | | | | | |
|---|---|---|---|---|---|---|
| Year | Rating | Default Rate | Calibrated Mean | Calibrated Median | 90% Confidence Interval | |
| | | | | | LB | UB |
| 2016 | 1 | 0.00% | 0.05% | 0.05% | 0.03% | 0.06% |
| | 2 | 0.00% | 0.09% | 0.08% | 0.08% | 0.10% |
| | 3 | 0.00% | 0.13% | 0.13% | 0.13% | 0.14% |
| | 4 | 0.00% | 0.22% | 0.22% | 0.20% | 0.24% |
| | 5 | 0.00% | 3.00% | 3.00% | 2.96% | 3.06% |
| | 6 | 0.08% | 3.48% | 3.48% | 3.44% | 3.55% |
| | 7 | 2.75% | 10.77% | 10.77% | 10.61% | 10.94% |
| | 8 | 0.00% | 15.64% | 15.67% | 15.15% | 16.13% |
| 2017 | 1 | 0.00% | 0.04% | 0.04% | 0.04% | 0.05% |
| | 2 | 0.00% | 0.08% | 0.08% | 0.07% | 0.08% |
| | 3 | 0.00% | 0.11% | 0.11% | 0.11% | 0.12% |
| | 4 | 0.00% | 0.16% | 0.16% | 0.15% | 0.17% |
| | 5 | 0.00% | 0.27% | 0.27% | 0.26% | 0.29% |
| | 6 | 0.08% | 0.57% | 0.57% | 0.56% | 0.59% |
| | 7 | 2.75% | 8.47% | 8.48% | 8.38% | 8.64% |
| | 8 | 0.00% | 18.87% | 19.05% | 18.18% | 19.57% |

---

[3] This part of the process was performed using R language which is a popular data science tool used in the industry.

*Figure 1: Calibrated PDs vs observed default rates for 2016 and 2017, n=100,000 simulations, K = 300 iterations*

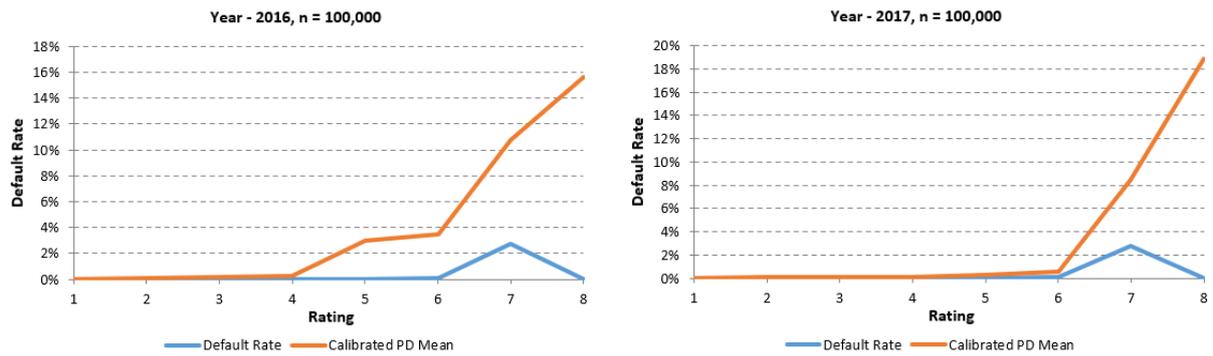

The distribution of calibrated PDs for all the ratings from 1 to 8 for both the years have been provided in the Appendix B. It is interesting to note that for some ratings, the distribution shows gaussian-like densities while for some others, the distribution shows 2 or more peaks, indicating mixture of multiple gaussian-like densities.

*Table 4: Calibrated PD values for 500,000 simulations and 300 iterations*

| Simulated Results n = 500,000 K =300 | | | | | | |
|---|---|---|---|---|---|---|
| Year | Rating | Default Rate | Calibrated Mean | Calibrated Median | 90% Confidence Interval | |
| | | | | | LB | UB |
| 2016 | 1 | 0.00% | 0.05% | 0.06% | 0.03% | 0.06% |
| | 2 | 0.00% | 0.08% | 0.08% | 0.08% | 0.09% |
| | 3 | 0.00% | 0.13% | 0.13% | 0.13% | 0.14% |
| | 4 | 0.00% | 0.21% | 0.21% | 0.20% | 0.23% |
| | 5 | 0.00% | 3.00% | 3.00% | 2.96% | 3.03% |
| | 6 | 0.08% | 3.48% | 3.48% | 3.44% | 3.51% |
| | 7 | 2.75% | 10.79% | 10.80% | 10.61% | 10.97% |
| | 8 | 0.00% | 15.71% | 15.70% | 15.15% | 16.13% |
| 2017 | 1 | 0.00% | 0.05% | 0.05% | 0.04% | 0.05% |
| | 2 | 0.00% | 0.08% | 0.08% | 0.07% | 0.08% |
| | 3 | 0.00% | 0.11% | 0.11% | 0.11% | 0.11% |
| | 4 | 0.00% | 0.16% | 0.16% | 0.15% | 0.17% |
| | 5 | 0.00% | 0.27% | 0.27% | 0.26% | 0.29% |
| | 6 | 0.08% | 0.57% | 0.57% | 0.56% | 0.58% |
| | 7 | 2.75% | 8.50% | 8.51% | 8.38% | 8.63% |
| | 8 | 0.00% | 18.76% | 18.60% | 18.18% | 19.51% |

*Figure 2: Calibrated PDs vs observed default rates for 2016 and 2017, n = 500,000 simulations, K = 300 iterations*

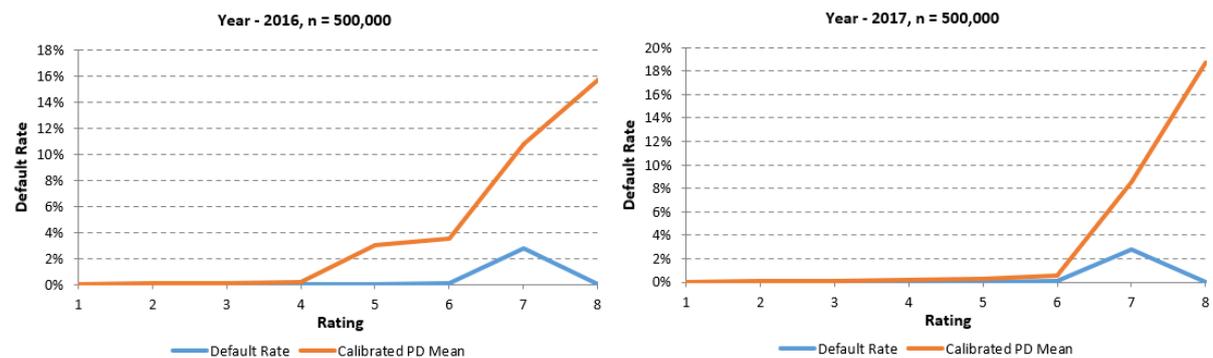

# 4. Comparison with other methods

The results obtained was also compared with other popular calibration techniques in the industry.

1. The *most prudent estimation* PDs suggested by Pluto & Tasche (2005).
2. *CAP Curve Calibration* or *VDB Calibration* suggested by Van der Burgt (2007)
3. *Quasi Moment Matching* (QMM) proposed by Dirk Tasche (2009).

In order to compare the models, all PD values were scaled to their corresponding year's central tendency or Portfolio average default rate. This is done so that the estimated PD at portfolio level is same across all the methods.

*Table 5: Portfolio distribution by year*

| Year | Customers | Defaults | Average Default Rate |
|---|---|---|---|
| 2016 | 124 | 5968 | 2.08% |
| 2017 | 51 | 7773 | 0.66% |

$$\frac{\overline{\theta_1} n_1 + \overline{\theta_2} n_2 + \cdots + \overline{\theta_m} n_m}{n_1 + n_2 + \cdots + n_m} = \theta_{Port} \tag{14}$$

Where $\overline{\theta_i}$ is the scaled default rate and $n_i$ is the number of observations of rating i, respectively. $\theta_{Port}$ is the portfolio default rate.

The values after scaling the PD values from each calibration technique, have been displayed in the Table 6 given below. From the table, we can observe a significant variation among the different methods in the worst rating grades. A visual comparison of PD trend can be observed from Figure 3 and Figure 4.

*Table 6: Comparison of PD results (scaled to average default rate) from different methods*

| Year | Rating | Simulated Mean (n = 100,000) | Simulated Mean (n = 500,000) | Pluto-Tasche (conf = 75%) | CAP Calibration | QMM |
|---|---|---|---|---|---|---|
| 2016 | 1 | 0.05% | 0.05% | 1.17% | 0.26% | 0.03% |
| | 2 | 0.08% | 0.08% | 1.17% | 0.26% | 0.09% |
| | 3 | 0.12% | 0.12% | 1.20% | 0.28% | 0.27% |
| | 4 | 0.21% | 0.20% | 1.43% | 0.52% | 0.71% |
| | 5 | 2.82% | 2.82% | 2.27% | 1.67% | 1.63% |
| | 6 | 3.27% | 3.27% | 2.33% | 4.29% | 3.68% |
| | 7 | 10.12% | 10.14% | 6.42% | 9.44% | 9.82% |
| | 8 | 14.70% | 14.76% | 6.42% | 11.67% | 24.89% |
| 2017 | 1 | 0.03% | 0.04% | 0.28% | 0.00% | 0.00% |
| | 2 | 0.07% | 0.07% | 0.28% | 0.00% | 0.00% |
| | 3 | 0.09% | 0.09% | 0.28% | 0.00% | 0.00% |
| | 4 | 0.13% | 0.13% | 0.32% | 0.00% | 0.03% |
| | 5 | 0.23% | 0.23% | 0.46% | 0.00% | 0.10% |
| | 6 | 0.48% | 0.48% | 0.66% | 0.00% | 0.52% |
| | 7 | 7.09% | 7.12% | 3.76% | 11.13% | 6.56% |
| | 8 | 15.80% | 15.71% | 8.16% | 33.96% | 38.43% |

*Figure 3: Comparison of results with estimates obtained from other methods for year 2016*

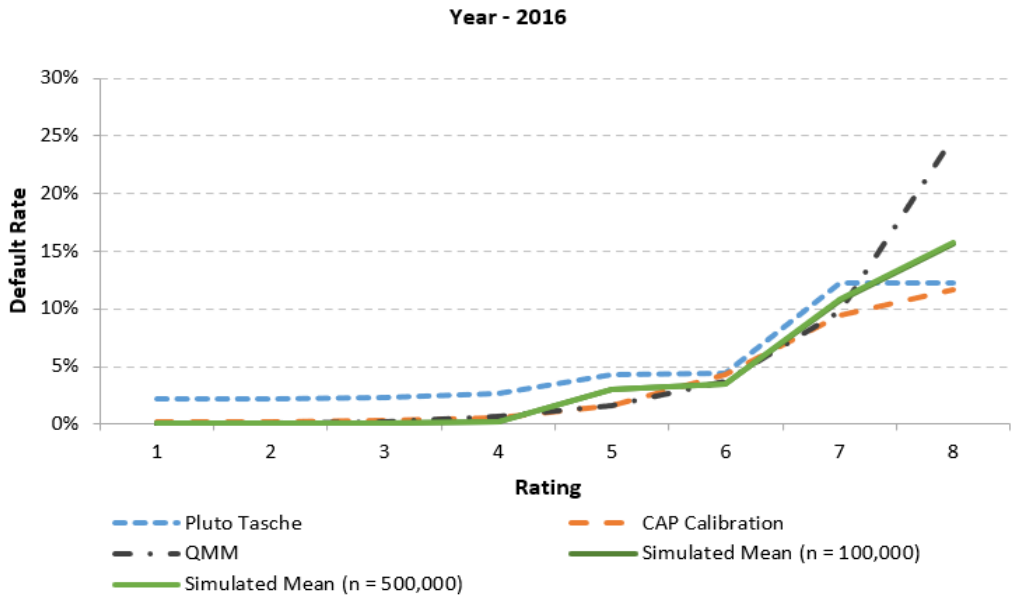

*Figure 4: Comparison of results with estimates obtained from other methods for year 2017*

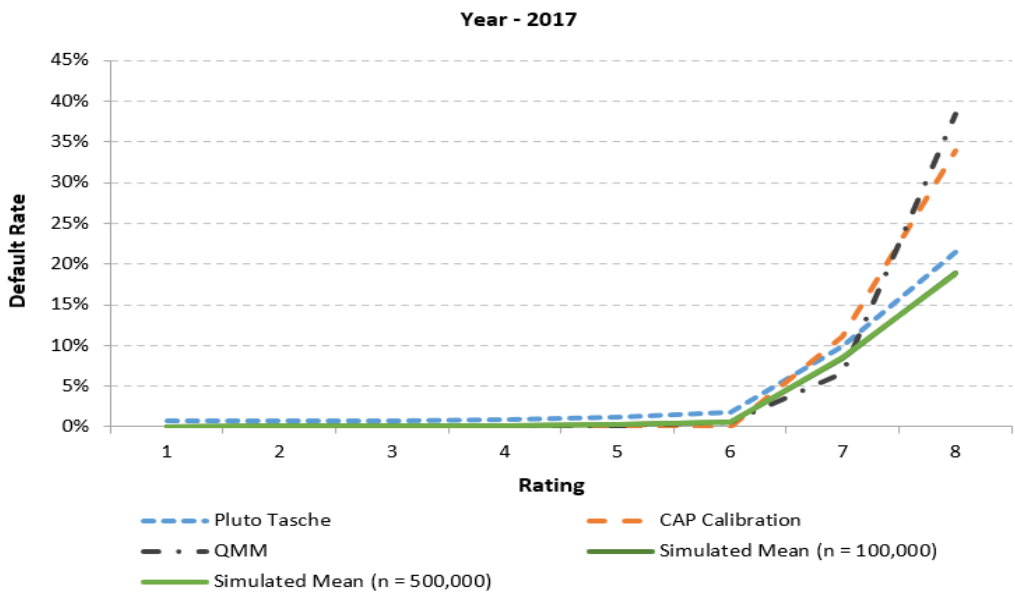

- The comparison exercise performed using *Pluto-Tasche Calibration*, *QMM* and *CAP calibration method* suggests that simulation-based approach provides reasonable results in line other industry-best practise techniques. The values are neither under-predicting nor over-predicting in comparison to the other models.
- It can be observed that QMM method provided significantly higher PD results in worse rating grades in both the years, while Pluto Tasche method generated slightly lower PDs when compared to all the others in the same range.
- It can also be observed that simulation-based results give satisfactorily close results irrespective of the number of simulations (n = 100,000 vs n = 500,000). Even in rating grade 8, where the separation is maximum, the difference between the two PDs is just 15.80% - 15.71% = 0.09%. This is advantageous as we need not choose large sampling size for the simulation of results. But even with 100,000 simulations, the algorithm consumed significant amount of time. For higher precision results, more simulations may be required which may be limited by the computational power of the user's computer system.

## 5. Potential application - Regression with exogenous variables.

So far, we have only considered the case where estimations are carried out on historical one-year observation periods for which data is available. But in many occasions, we may be required to estimate PDs for different points in time when default data is either unavailable (such as years before the start of default information) or unobserved (such as future time periods). In order to build a PD term structure (default probabilities for multiple future time periods), we need a robust method for establishing a relationship between PD values and exogenous regressor variables (for example systemic factors like GDP growth or Unemployment Rate). One generally accepted method of estimation is to use the PD from the historical time periods to build a regression model where the PD values or a transformed version of PD values are modelled as a linear function of one or more macro-economic factors or idiosyncratic factors.

$$\Phi(\theta_R) = \beta_0 + \beta_1 y_1 + \beta_2 y_2 + \cdots + \beta_k y_k \tag{15}$$

Our goal here, is to propose a regression model that is tailored for situations where the dependent variable ($\theta$) is measured continuously on the standard unit interval, i.e. $0 < \theta_R < 1$. Beta regression model is based on the assumption that the response is beta distributed. Since our estimated means are from calibrated beta distributions, the beta regression model fits perfectly into the PD modelling framework. Ferrari and Cribari-Neto (2004) first proposed the beta regression model for continuous variates that assume values in the standard unit interval, e.g., rates, proportions, or concentration indices. Hence, we do not present the methodology in full generality with example but rather introduce it as a concept.

Let y1, y2…., yk be independent random variables, then the beta regression model assumes that if $\theta_R \sim Beta(\alpha, \beta)$ then,

$$logit^{-1}(\psi) = \mu_\theta = \frac{\alpha}{\alpha + \beta} \tag{16}$$

$$\psi = \beta_0 + \beta_1 y_1 + \beta_2 y_2 + \cdots + \beta_k y_k \tag{17}$$

$logit^{-1}(\psi)$ can be replaced with any monotonically increasing link function. $\{\beta_1, \dots, \beta_k\}$ are the regression coefficients of each independent variable and $\beta_0$ is the intercept.

Using the estimated values of Probability of Default $\hat{\theta}_R(t)$ for historical snapshot periods from the previous approach and exogenous predictor variables $(y_1(t), y_2(t) \dots y_k(t))$ we can train a model for beta regression parameters $(\hat{\beta}_0, \hat{\beta}_1, \hat{\beta}_2, \dots, \hat{\beta}_k)$. The model can then be used to predict the value of mean $\mu^*$ given new data point $(y_1^*, y_2^*, \dots, y_k^*)$.

One main advantage of our proposed PD estimation method is that, since the default rate parameter $\theta_R$, is assumed to be a beta random variable, it naturally incorporates into the beta regression framework. Another advantage is that since the historical PD values are already calibrated in the 1st step, the estimated future PDs from regression model will also be calibrated in nature.

## 6. Conclusion

In this paper, we have presented a method for estimation and calibration of rating systems, which solves the issue of non-monotonicity which creeps in when default rates are empirically determined. The method introduced in the document is implemented by assuming that the default rate parameter is a random variable that follows Beta distribution. A simulation algorithm is designed to sample from the conditional distribution of default rate parameter for each rating. In this paper, the method is demonstrated for a portfolio of corporate entities rated by Standard and Poor's credit rating agency and the results obtained have been observed to be comparable to that of other popular calibration methods that are widely used.

One advantage of this method is that it is conceptually simple and straight forward and has minimum number of assumptions. The simulation algorithm is also simple and easy to implement using a statistical software. The results can be scaled to any appropriate central tendency. It has also been elucidated, how this method can be used for estimating probability of default for future point in times and how it integrates well with a Beta regression framework.

Although this approach is simple and straightforward, one drawback is that the process consumes significant amount of time. For higher precision and for large number of ratings, more simulations may be required which may be limited by the computational power of the user's computer system. Further researches need to be conducted on how to improve the framework by using faster sampling with improved accuracy of the PD estimates.

## References

Basel Committee on Banking Supervision (2001). The Internal Ratings-Based Approach Consultative Document. https://www.bis.org/publ/bcbsca05.pdf

Krahnen, J.P & Weber, M. Generally accepted rating principles: A primer. *Journal of Banking & Finance*, Elsevier, January, vol 25(1): 3-23

S&P (2020). Default, Transition, and Recovery: 2020 Annual Global Corporate Default And Rating Transition Study. https://www.maalot.co.il/Publications/TS20210408160139.PDF

Dirk Tasche, (2009) Estimating discriminatory power and PD curves when the number of defaults is small. *Working paper, Lloyds Banking Group*.

Pluto, K. and Dirk Tasche, (2005). Thinking Positively. *Risk*, August, 72-78

Dirk Tasche, (2013) The art of probability-of-default curve calibration. *Journal of Credit Risk*, 9:63-103.

Dirk Tasche, (2002) Remarks on the monotonicity of default probabilities

Van der Burgt, M. (2008) Calibrating low-default portfolios, using the cumulative accuracy profile. *Journal of Risk Model Validation*, 1(4):17-33.

Ferrari SLP, Cribari-Neto F (2004). "Beta Regression for Modelling Rates and Proportions." *Journal of Applied Statistics*, 31(7), 799–815.

# Appendix A

**Method of Moments: Beta Distribution**

Given a sample of data that may fit the beta distribution, we want to estimate the parameters which best fit the data. We can estimate the sample mean for the beta distribution by the population mean, as follows:

$$\tilde{x} \approx \frac{\alpha}{\alpha + \beta} \tag{A.118}$$

We treat this as equation and solve for $\alpha$ and $\beta$. From the first equation, we get

$$\beta \approx \frac{\alpha(1 - \tilde{x})}{\tilde{x}} \tag{A.219}$$

Similarly, we can estimate the sample variance from the data and relate it to the population variance, as follows:

$$s^2 \approx \frac{\alpha\beta}{(\alpha + \beta)^2(\alpha + \beta + 1)} \tag{A.320}$$

$$s^2 \approx \frac{\tilde{x}\beta}{(\alpha + \beta)(\alpha + \beta + 1)} = \frac{\tilde{x}}{\left(\alpha + \frac{\alpha(1 - \tilde{x})}{\tilde{x}}\right)\left(\alpha + \frac{\alpha(1 - \tilde{x})}{\tilde{x}} + 1\right)} \frac{\alpha(1 - \tilde{x})}{\tilde{x}} \tag{A.421}$$

$$s^2 \approx \frac{\alpha(1 - \tilde{x})\tilde{x}^2}{\left(\alpha\tilde{x} + \alpha(1 - \tilde{x})\right)(\alpha\tilde{x} + \alpha(1 - \tilde{x}) + \tilde{x})} \tag{A.522}$$

$$s^2 \approx \frac{\alpha(1 - \tilde{x})\tilde{x}^2}{(\alpha)(\alpha + \tilde{x})} = \frac{(1 - \tilde{x})\tilde{x}^2}{(\alpha + \tilde{x})} \tag{A.623}$$

$$\alpha \approx \frac{(1 - \tilde{x})\tilde{x}^2}{s^2} - \tilde{x} = \tilde{x}\left(\frac{\tilde{x}(1 - \tilde{x})}{s^2} - 1\right) \tag{A.724}$$

# Appendix B

**Distribution of sampled values, n = 100,000, year = 2016**

*Figure 5: Distribution of simulated PDs Rating 1, n=100,000, year = 2016*

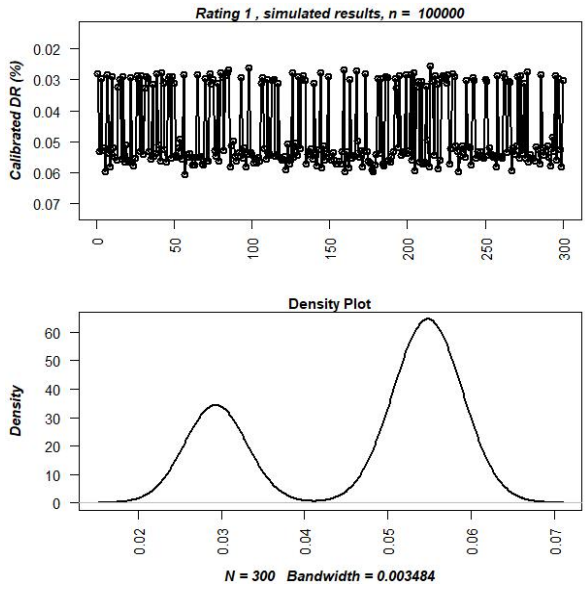

*Figure 6: Distribution of simulated PDs Rating 2, n=100,000, year = 2016*

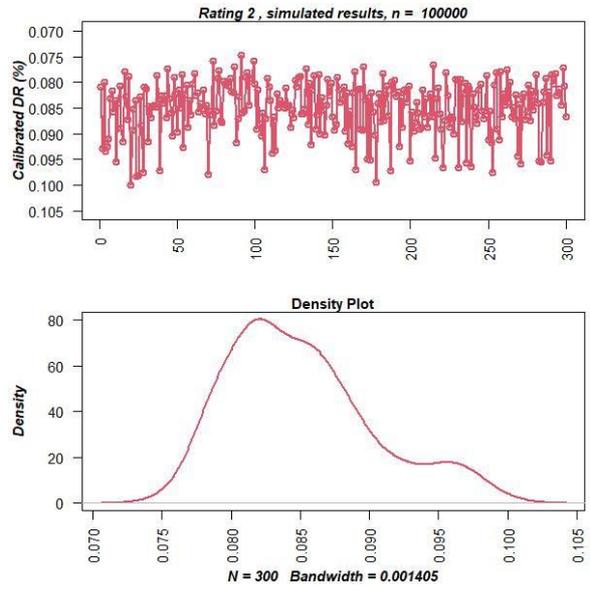

*Figure 7: Distribution of simulated PDs Rating 3, n=100,000, year = 2016*

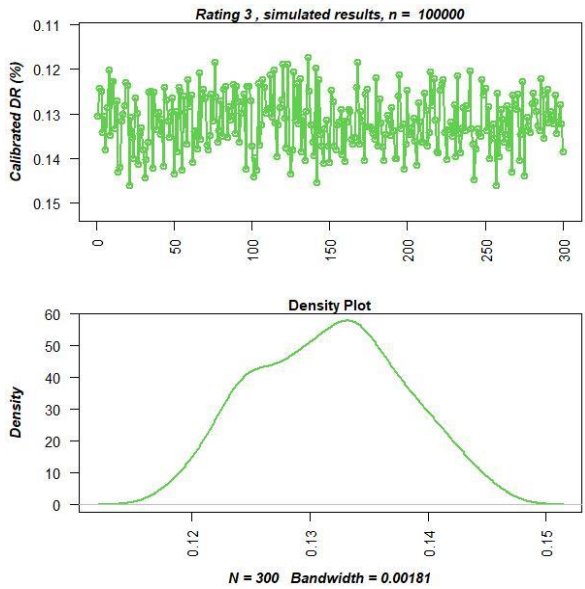

*Figure 8: Distribution of simulated PDs Rating 4, n=100,000, year = 2016*

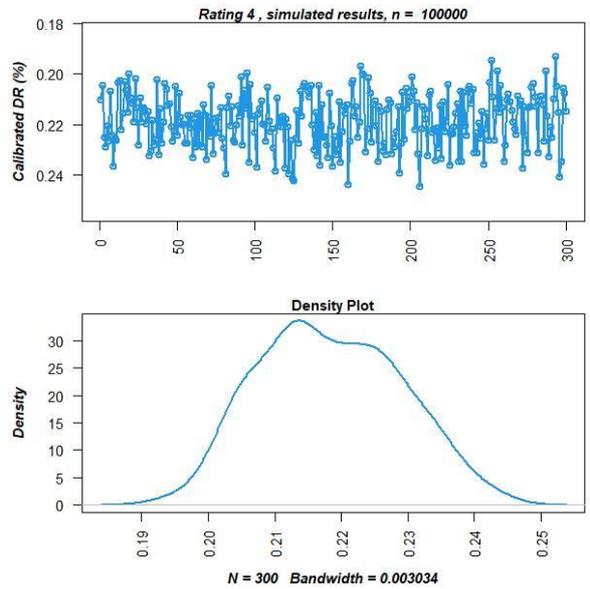

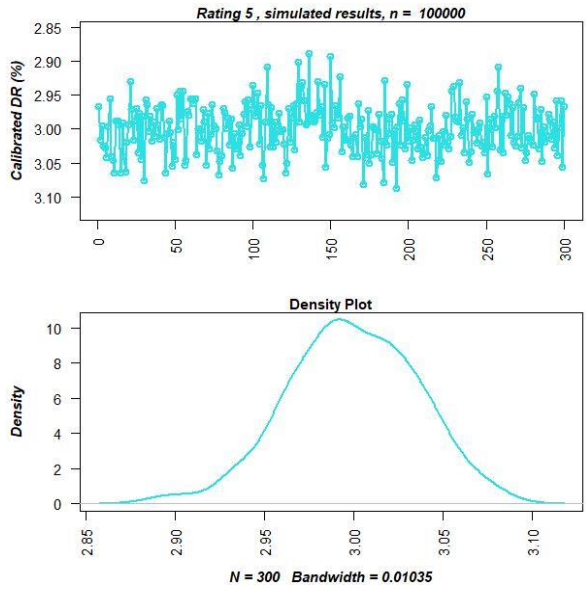

*Figure 9: Distribution of simulated PDs Rating 5, n=100,000, year = 2016*

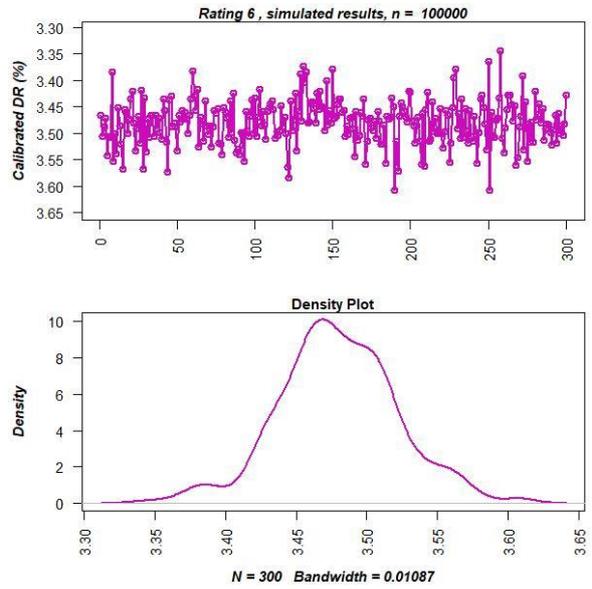

*Figure 10: Distribution of simulated PDs Rating 6, n=100,000, year = 2016*

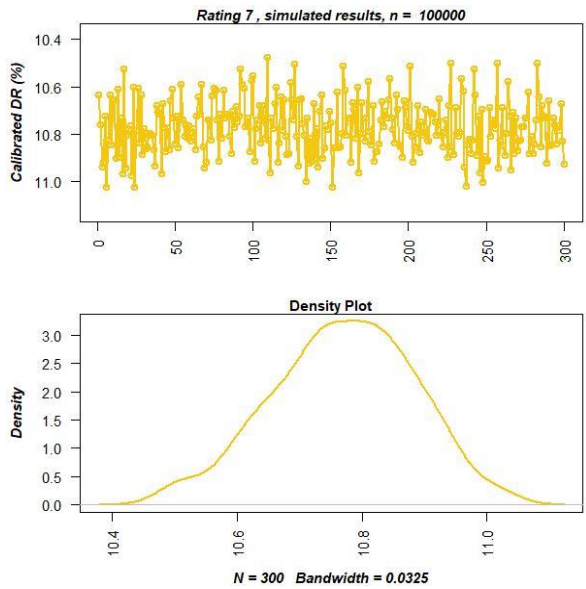

*Figure 11: Distribution of simulated PDs Rating 7, n=100,000, year = 2016*

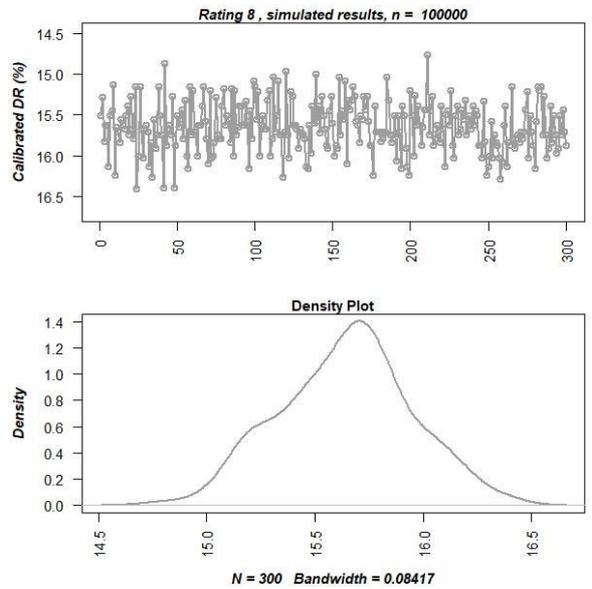

*Figure 12: Distribution of simulated PDs Rating 8, n=100,000, year = 2016*

## Distribution of sampled values, n = 500,000, year = 2016

*Figure 13: Distribution of simulated PDs Rating 1, n=500,000, year = 2016*

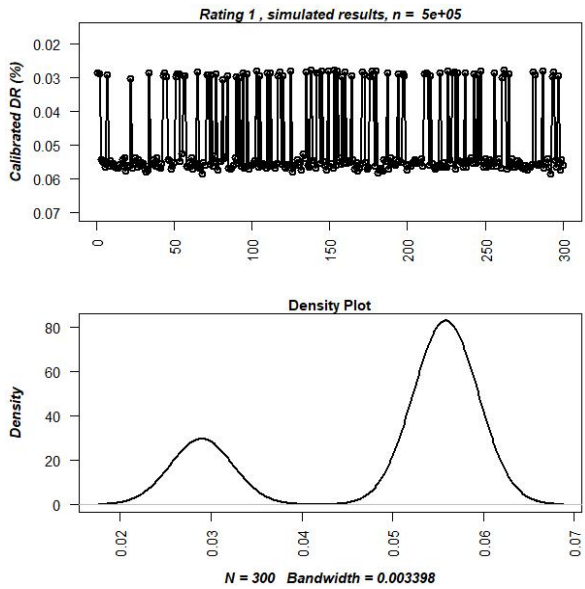

*Figure 14: Distribution of simulated PDs Rating 2, n=500,000, year = 2016*

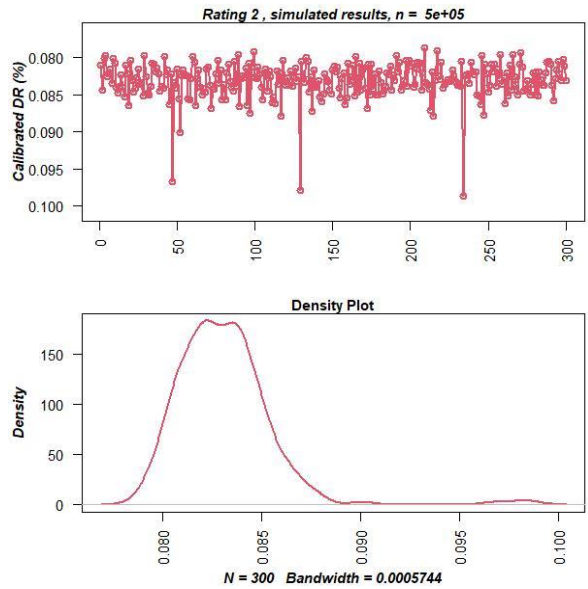

*Figure 15: Distribution of simulated PDs Rating 3, n=500,000, year = 2016*

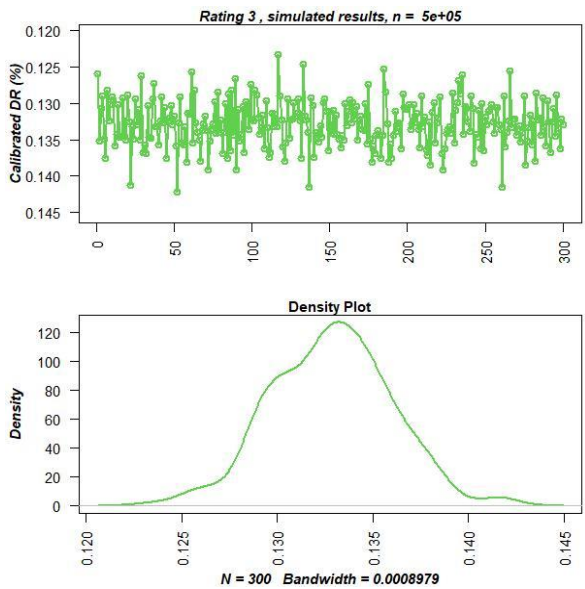

*Figure 16: Distribution of simulated PDs Rating 4, n=500,000, year = 2016*

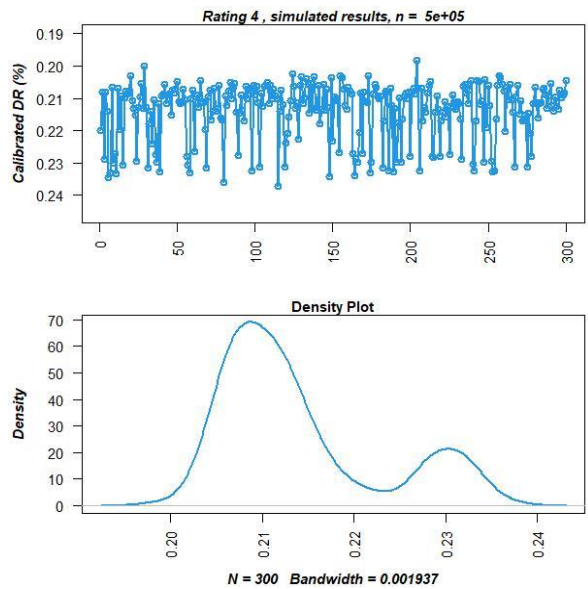

*Figure 17: Distribution of simulated PDs Rating 5, n=500,000, year = 2016*

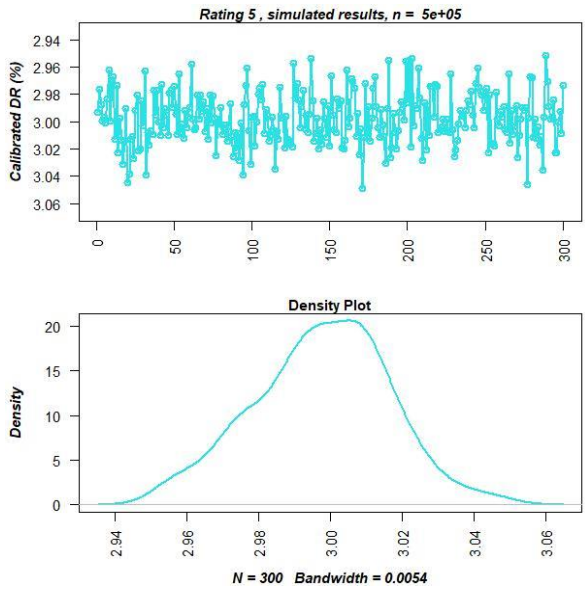

*Figure 18: Distribution of simulated PDs Rating 6, n=500,000, year = 2016*

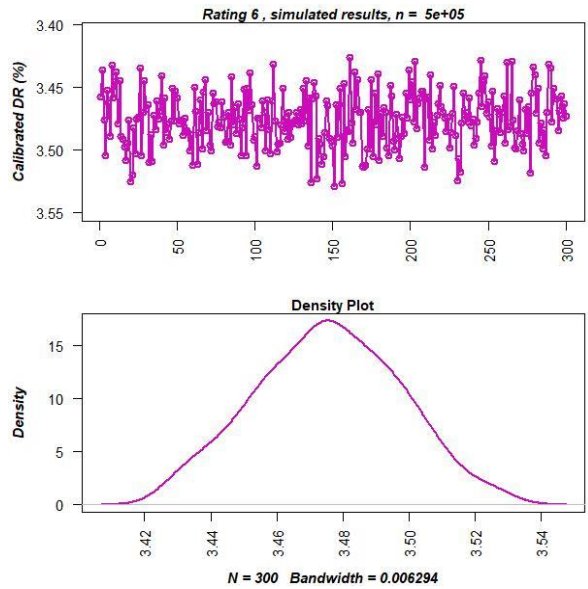

*Figure 19: Distribution of simulated PDs Rating 7, n=500,000, year = 2016*

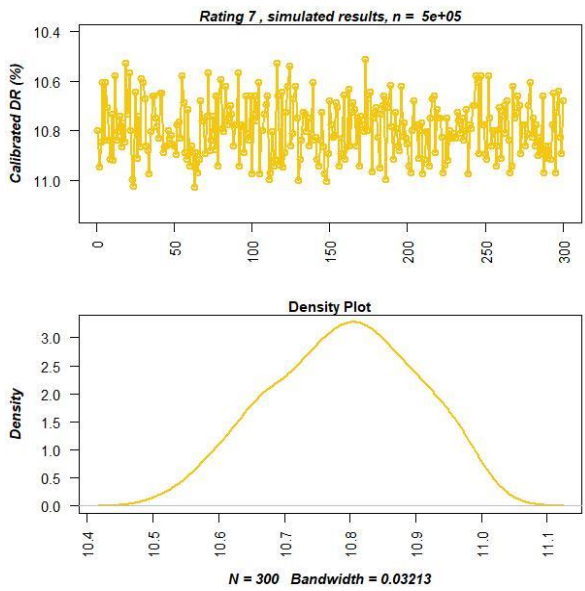

*Figure 20: Distribution of simulated PDs Rating 8, n=500,000, year = 2016*

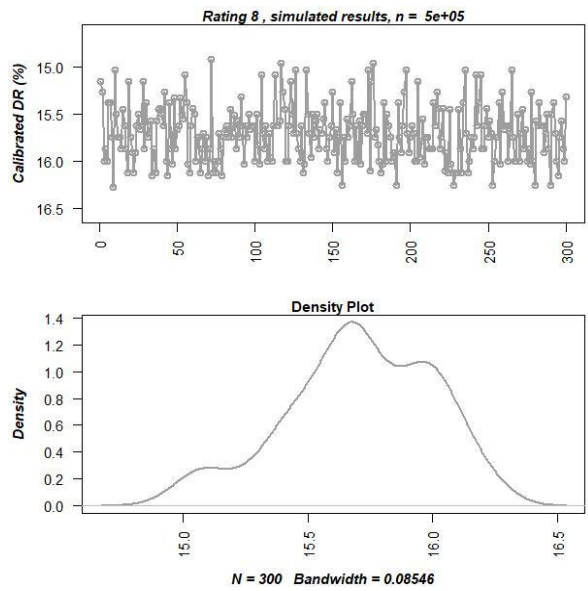

**Distribution of sampled values, n = 100,000, year = 2017**

*Figure 21: Distribution of simulated PDs Rating 1, n=100,000, year = 2017*

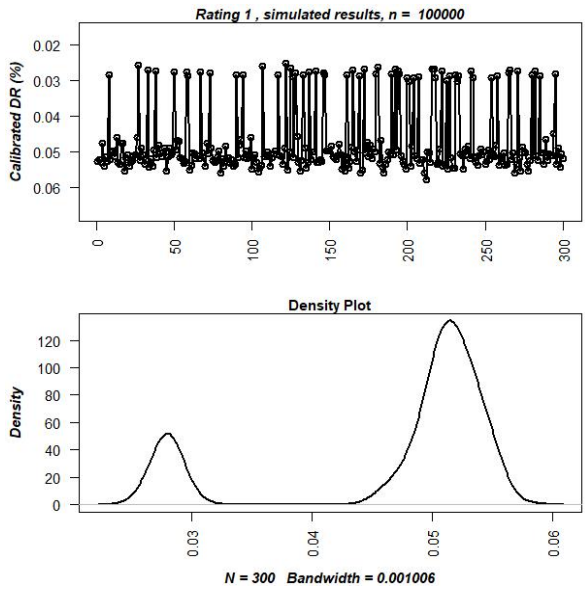

*Figure 22: Distribution of simulated PDs Rating 2, n=100,000, year = 2017*

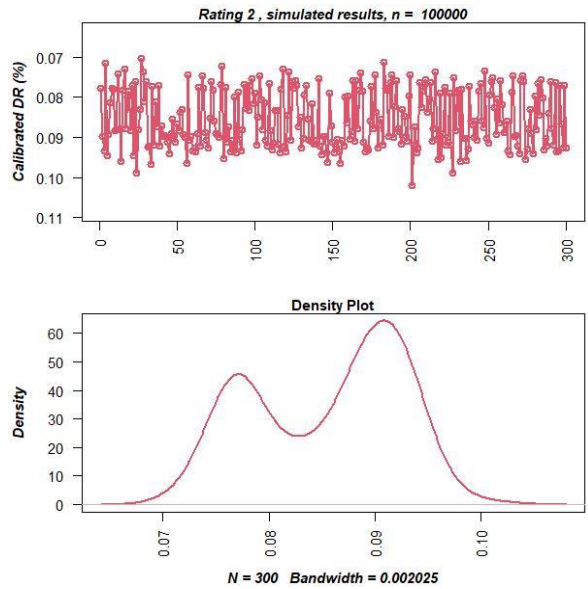

*Figure 23: Distribution of simulated PDs Rating 3, n=100,000, year = 2017*

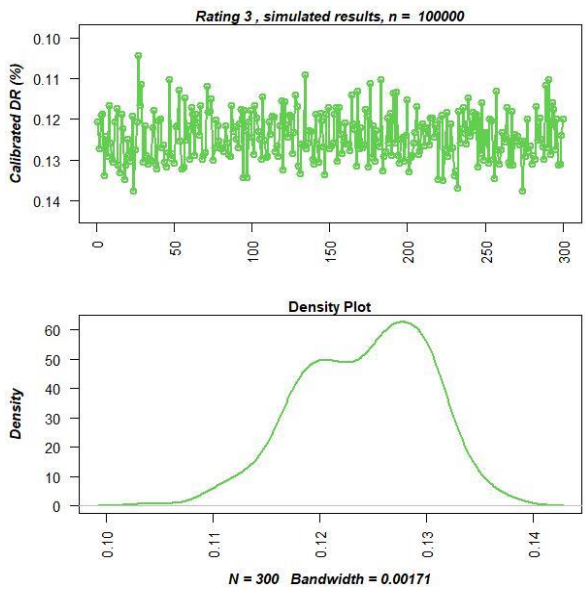

*Figure 24: Distribution of simulated PDs Rating 4, n=100,000, year = 2017*

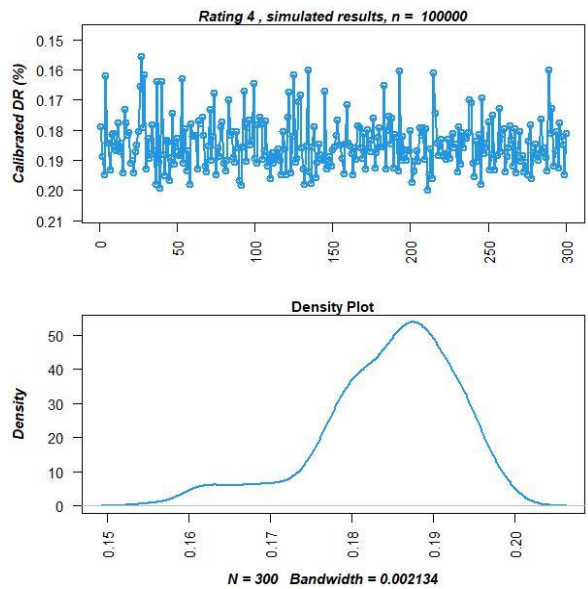

*Figure 25: Distribution of simulated PDs Rating 5, n=100,000, year = 2017*

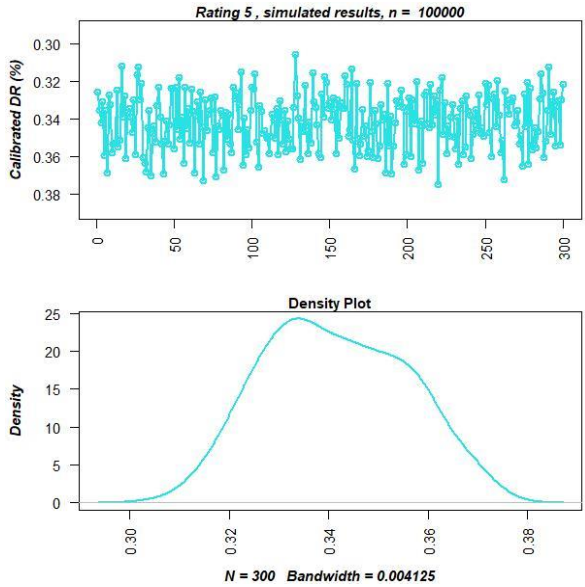

*Figure 26: Distribution of simulated PDs Rating 6, n=100,000, year = 2017*

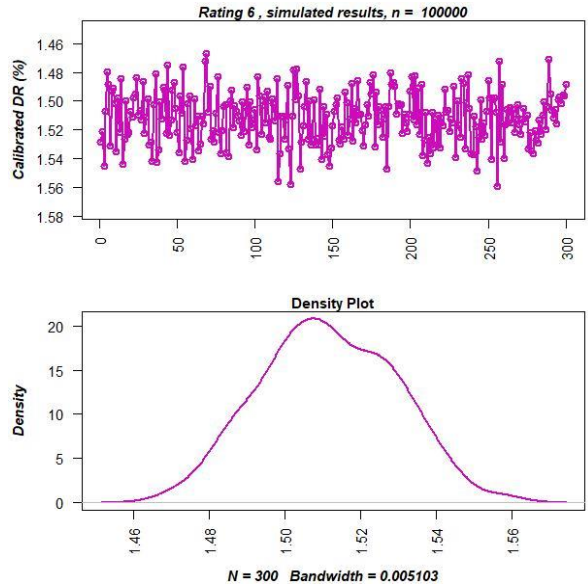

*Figure 27: Distribution of simulated PDs Rating 7, n=100,000, year = 2017*

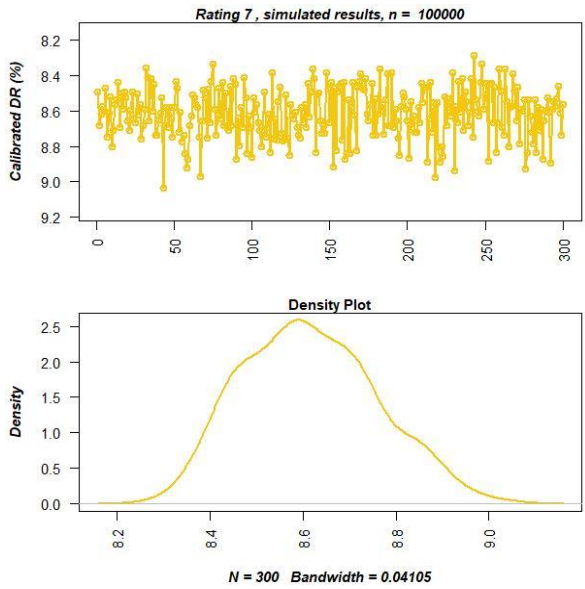

*Figure 28: Distribution of simulated PDs Rating 8, n=100,000, year = 2017*

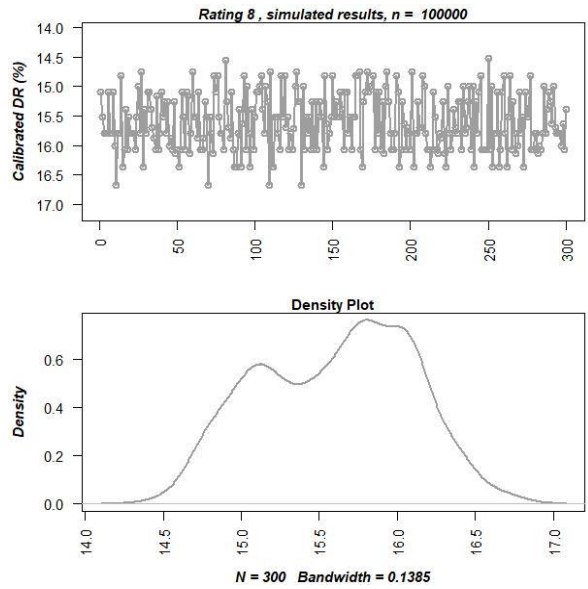

**Distribution of sampled values, n = 500,000, year = 2017**

*Figure 29: Distribution of simulated PDs Rating 1, n=500,000, year = 2017*

*Figure 30: Distribution of simulated PDs Rating 2, n=500,000, year = 2017*

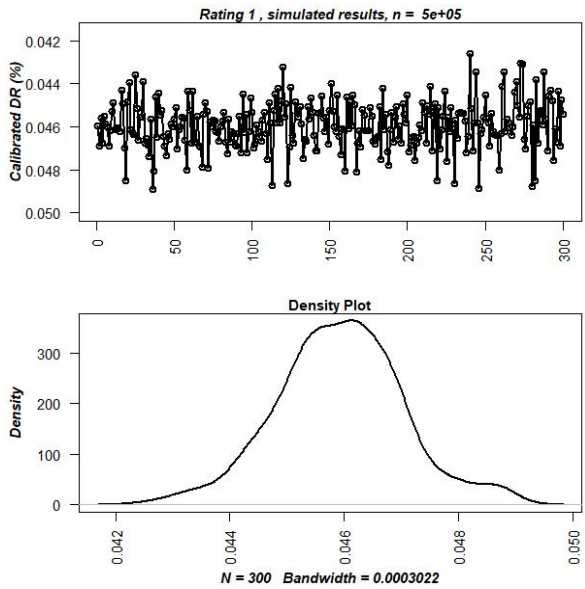
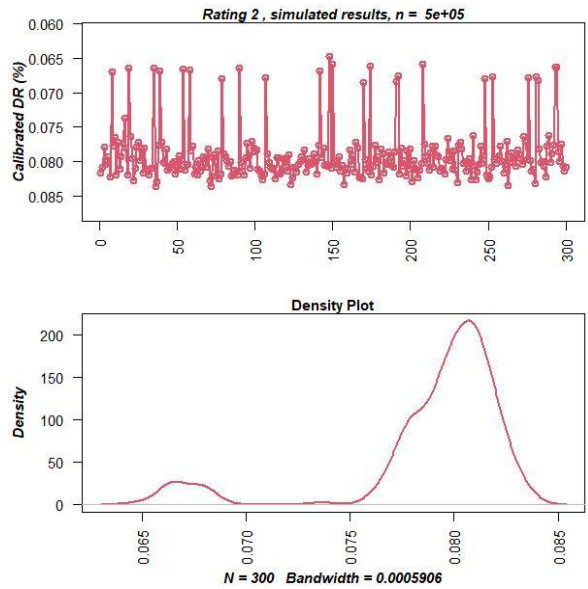

*Figure 31: Distribution of simulated PDs Rating 3, n=500,000, year = 2017*

*Figure 32: Distribution of simulated PDs Rating 4, n=500,000, year = 2017*

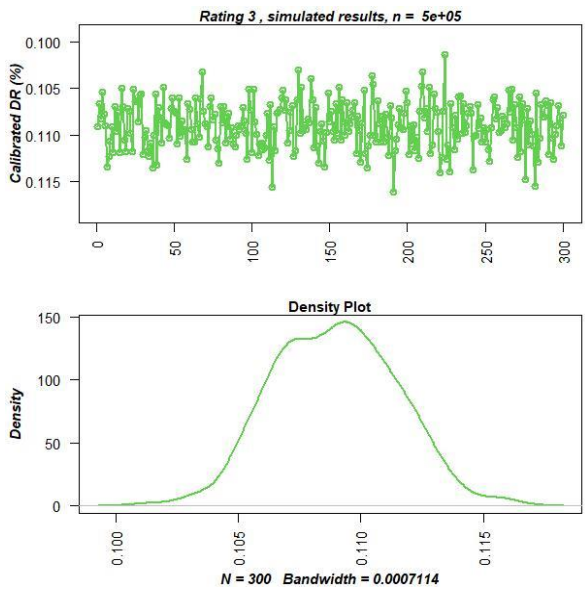
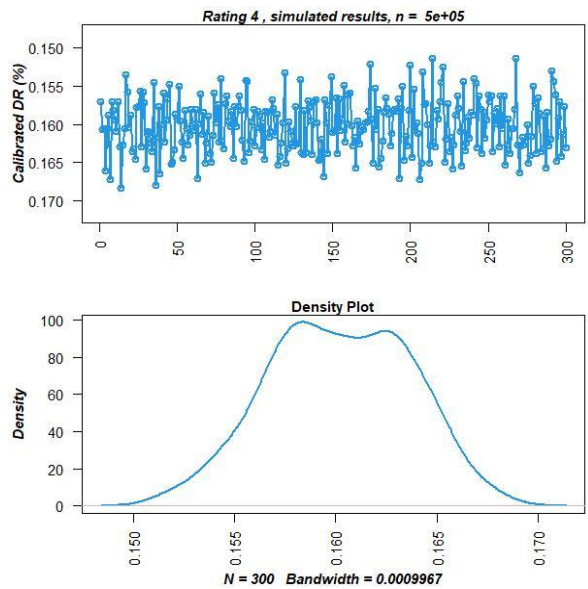

*Figure 33: Distribution of simulated PDs Rating 5, n=500,000, year = 2017*

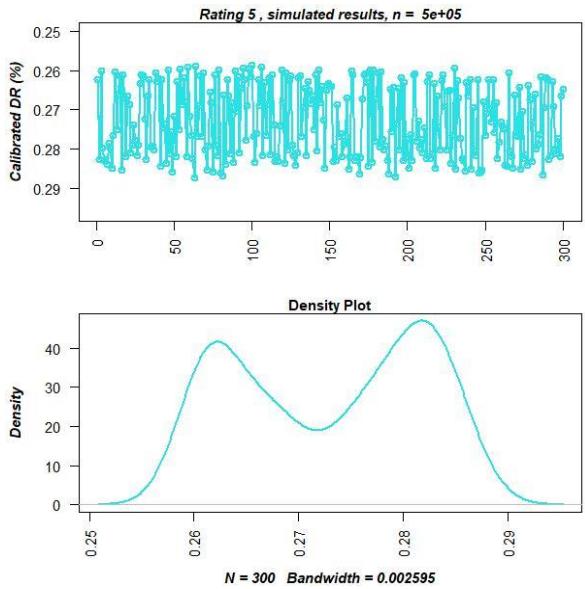

*Figure 34: Distribution of simulated PDs Rating 6, n=500,000, year = 2017*

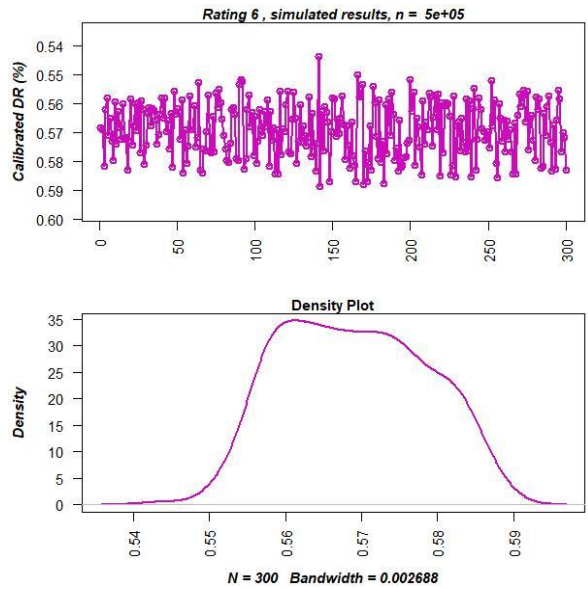

*Figure 35: Distribution of simulated PDs Rating 7, n=500,000, year = 2017*

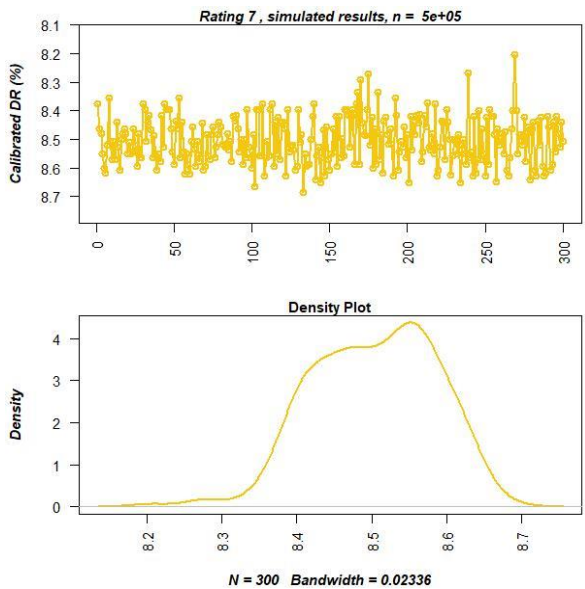

*Figure 36: Distribution of simulated PDs Rating 8, n=500,000, year = 2017*

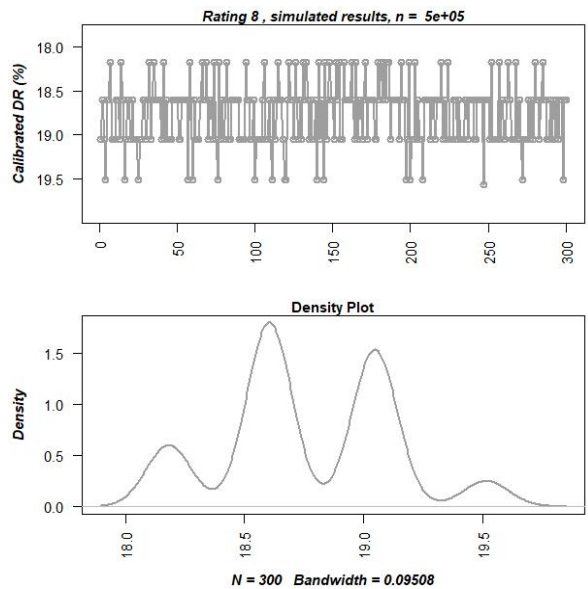